\documentclass[prl,twocolumn,floatfix,reprint]{revtex4}
\usepackage{graphicx,amssymb,amsmath}

\usepackage{graphicx}
\usepackage{dcolumn}
\usepackage{bm}
\usepackage{hyperref}
\usepackage{xcolor}

\definecolor{crimson}{HTML}{DC143C}
\definecolor{perfect_green}{HTML}{4FBF26}

\def\equationautorefname~#1\null{Eq.~(#1)\null}
\def\figureautorefname~#1\null{Fig.~#1\null}
\def\sectionautorefname~#1\null{Sec.~#1\null}

\begin{document}

\title{From lines to networks}

 \author{Marc Barthelemy$^{1,2}$}
 \affiliation{$^1$Université Paris-Saclay, CNRS, CEA, Institut de
   Physique Théorique, 91191 Gif-sur-Yvette, France}
 \affiliation{$^2$Centre d'Analyse et de Math\'ematique Sociales (CNRS/EHESS) 54 Avenue de Raspail, 75006 Paris, France}

\date{\today}

\begin{abstract}

Many real-world networks, ranging from subway systems to polymer structures and fungal mycelia, do not form by the incremental addition of individual nodes but instead grow through the successive extension and intersection of lines or filaments. Yet most existing models for spatial network formation focus on node-based growth, leaving a significant gap in our understanding of systems built from spatially extended components. Here we introduce a minimal model for spatial networks, rooted in the iterative growth and intersection of lines—a mechanism inspired by diverse systems including transportation networks, fungal hyphae, and vascular structures. Unlike classical approaches, our model constructs networks by sequentially adding lines across a domain populated with randomly distributed points. Each line grows greedily to maximize local coverage, while subject to angular continuity and the requirement to intersect existing structures. This emphasis on extended, interacting elements governed by local optimization and geometric constraints leads to the spontaneous emergence of a core-and-branches architecture. The resulting networks display a range of non-trivial scaling behaviors: the number of intersections grows subquadratically; Flory exponents and fractal dimensions emerge consistent with empirical observations; and spatial scaling exponents depend on the heterogeneity of the underlying point distribution, aligning with measurements from subway systems. Our model thus captures key organizational features observed across diverse real-world networks, establishing a universal paradigm that goes beyond node-based approaches and demonstrates how the growth of spatially extended elements can shape the large-scale architecture of complex systems.


\end{abstract}

\maketitle



\section{Introduction}

Over the past two decades, a large number of models have been proposed to explain the structure and growth of networks—spatial or not~\cite{latora2017complex,barthelemy2022spatial}. Most of these models rely on the sequential addition of nodes or links, typically connecting newly added nodes to an existing set according to heuristic rules such as preferential attachment, cost minimization, or spatial proximity. These approaches have proven successful in reproducing key structural properties observed in real-world networks, including scale-free degree distributions, small-world effects, and modular organization.

However, many real-world networks do not grow by simply adding nodes one at a time. Instead, their development often follows a fundamentally different process that is line-based rather than node-based. In such systems, the elementary construction unit is not a single node but an extended structure, such as a path or a line. Understanding the implications of this line-based growth mechanism is essential for characterizing the resulting topology and dynamics of these networks. Examples of line-based systems include polymer, fiber networks, or cross-linked semi flexible filaments~\cite{broedersz2014modeling,danielsen2021molecular,onck2005alternative}, fracture networks in disordered materials~\cite{sanderson2015use}, biological systems such as vascular or neural networks~\cite{manoussaki1996mechanical}, and infrastructure systems including roads, railways, and subway networks~\cite{Xie:2009,Derrible:2010b,Roth:2012,cats2017topological,barthelemy2022spatial}. Notably, the growth of filamentous fungi also follows a line-based pattern, as these organisms expand through apical hyphal extension and branching~\cite{edelstein1982propagation}, forming intricate mycelial networks optimized for foraging and nutrient transport~\cite{bebber2007biological,fricker2008interplay,jonathan2006reorganization,krull2015filaments,fricker2017}. Fungal networks are not generated from centralized plans or fixed templates but emerge through local, iterative developmental processes. Growth involves an initial overproduction of links and nodes, followed by selective pruning of some connections and reinforcement of others~\cite{fricker2009adaptive}.

\begin{figure}[htbp]
\centering
   \includegraphics[width=0.9\linewidth]{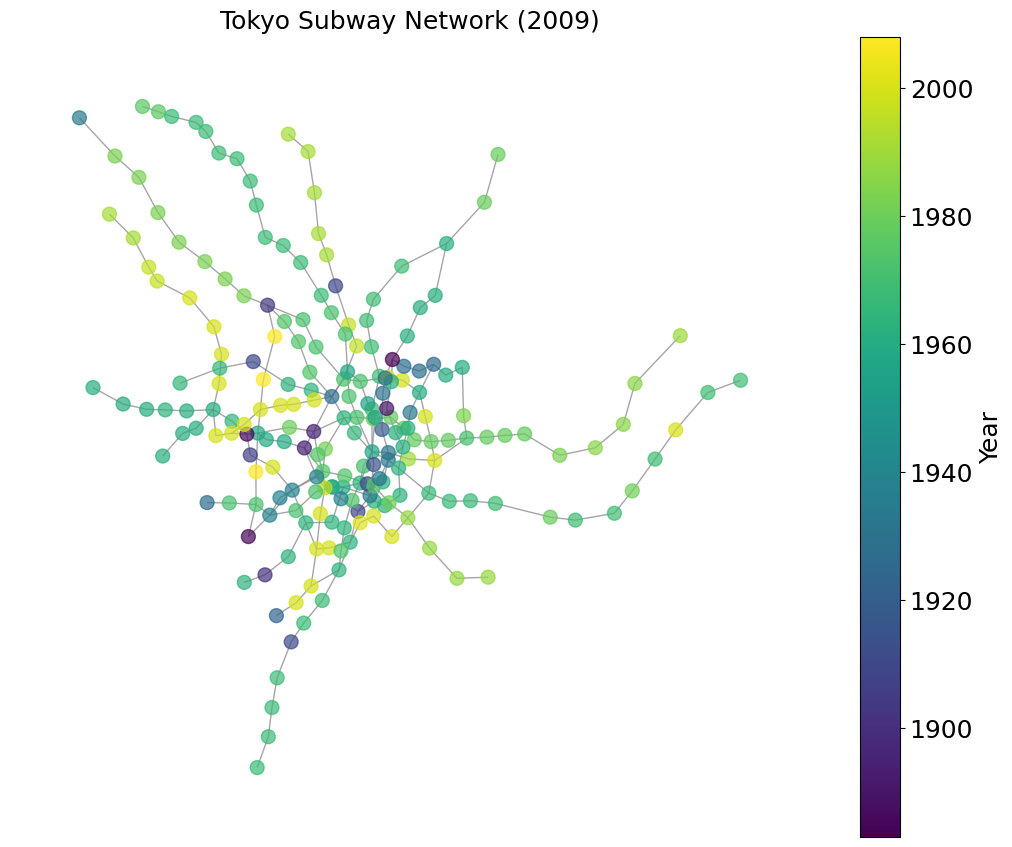} 
   \includegraphics[width=0.8\linewidth]{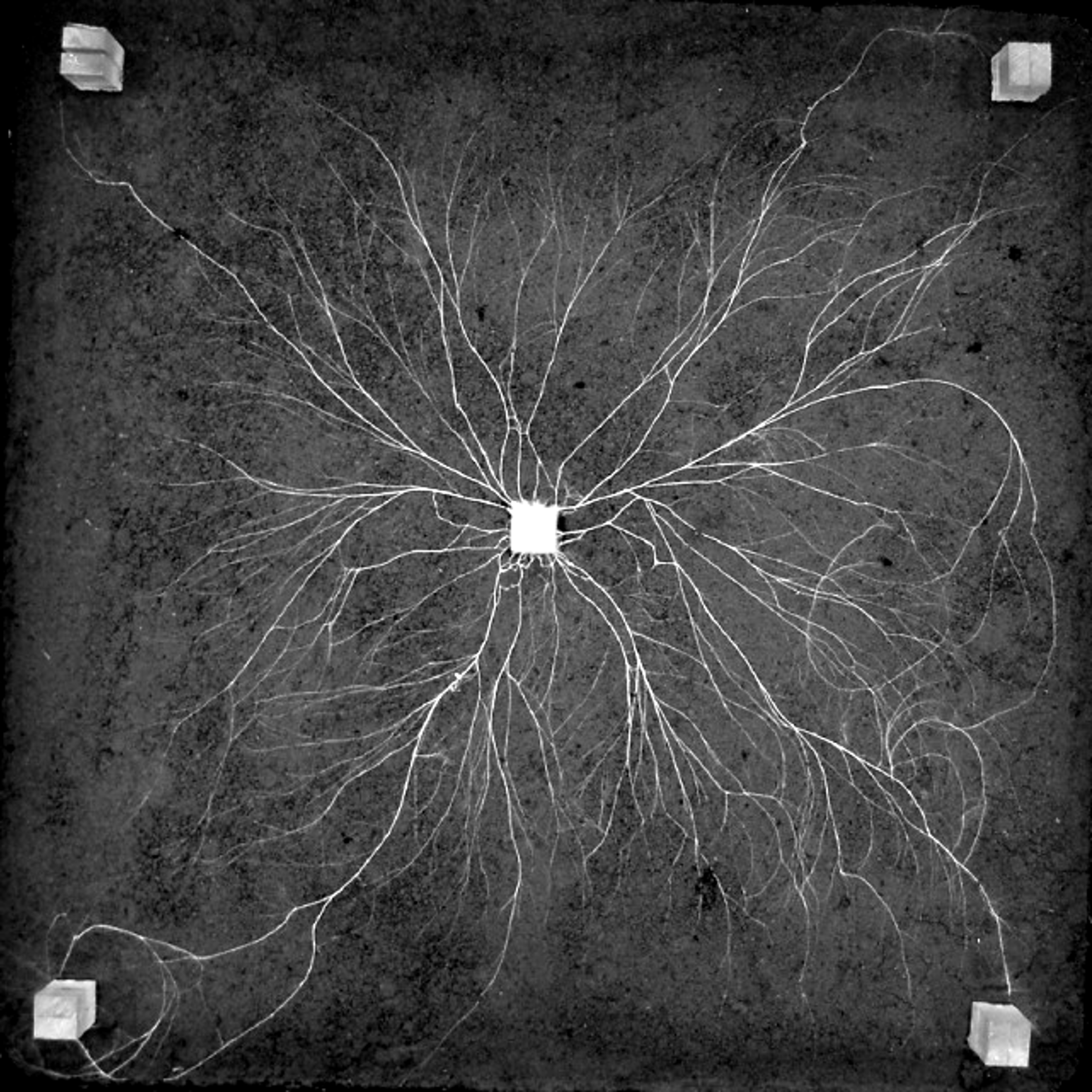} 
   \caption{Line-based systems. (a) Tokyo subway network (as of 2009). Colors indicate the year each station was opened. (b) Mycelial system growing from a central inoculum. Figure from \cite{jonathan2006reorganization} (with permission from the publisher).}
   \label{fig:tokyo}
\end{figure}

Among all these examples, transportation systems provide a particularly clear example of line-based growth: such networks typically expand through the successive addition of full routes or lines, rather than through incremental, link-by-link development. This line-based construction logic is especially salient in subway systems, where individual lines are designed and built at different stages of the network’s development, often with the objective of improving spatial coverage, system connectivity, and transport capacity \cite{laporte2019location}. Figure~\ref{fig:tokyo}(a) illustrates this with the case of the Tokyo subway network as of 2009, where the color of each station indicates its opening year. 

This figure underscores the fundamentally line-based nature of subway development: older stations and lines are concentrated near the center, while more recent expansions extend outward as peripheral branches. The continuity of colors along each line—indicating station opening years—suggests that transit lines are typically built as coherent units over relatively short time frames, rather than through incremental, station-by-station additions. These systems offer a rich empirical ground for examining how spatial and topological features emerge from the addition of linear infrastructure elements.

In Fig.~\ref{fig:tokyo}(b), we show a digital image of a mycelial network (from \cite{jonathan2006reorganization}), which similarly grows via line-like filaments radiating from a central origin. As reported in~\cite{jonathan2006reorganization}, this biological network exhibits branching, intersection, and merging dynamics that culminate in a spatial organization reminiscent of a core–branch structure. It has been argued that fungi grow as self-organized planar spatial networks shaped by evolutionary pressures~\cite{fricker2008interplay}, making them relevant for our study. Despite the distinct nature of the underlying processes—urban planning in one case, biological morphogenesis in the other—both systems illustrate how line-based growth, spatial constraints, and connectivity requirements can drive the emergence of similar topological patterns.

Despite the diversity of models for real-world networks, there remains a gap in frameworks that explicitly model network growth via the addition of lines. One notable exception is the model introduced in~\cite{devroye1994intersections,bottcher2020random}, in which line segments are randomly added in space, and their intersections define the nodes of the resulting graph. In this work, we propose a new model for generating spatial networks through the sequential addition of lines within a confined, disk-shaped area. Each added line aims to optimize coverage while intersecting existing lines to ensure network connectivity. This line-based growth process reproduces for example essential features observed in real-world systems and opens new perspectives for understanding spatial networks more generally. Although our comparison point here is mostly about empirical subway networks, the relevance of this model extends beyond urban transportation and provides a new paradigm for the study of spatially embedded networks.

\section{The model}

We introduce a minimal spatial model for the growth of networks constructed through the successive addition of lines, rather than isolated nodes. This generative mechanism mimics some of the processes observed in transportation systems \cite{laporte2019location}, polymer networks, or biological filaments, where spatial constraints and coverage efficiency play a critical role in shaping the network’s topology. 

We start with the spatial setup. The network grows within a circular domain of radius $R$. A set of $P$ points is sampled from an exponential radial density distribution
\begin{align}
\rho(r) = \rho_0 e^{-r/r_0},
\end{align}
where $r_0$ controls the decay length. These points represent the spatial demand (e.g., individuals in a transportation scenario), which the network aims to service efficiently.

\begin{figure}[htp]
    \centering
    \includegraphics[width=0.7\linewidth]{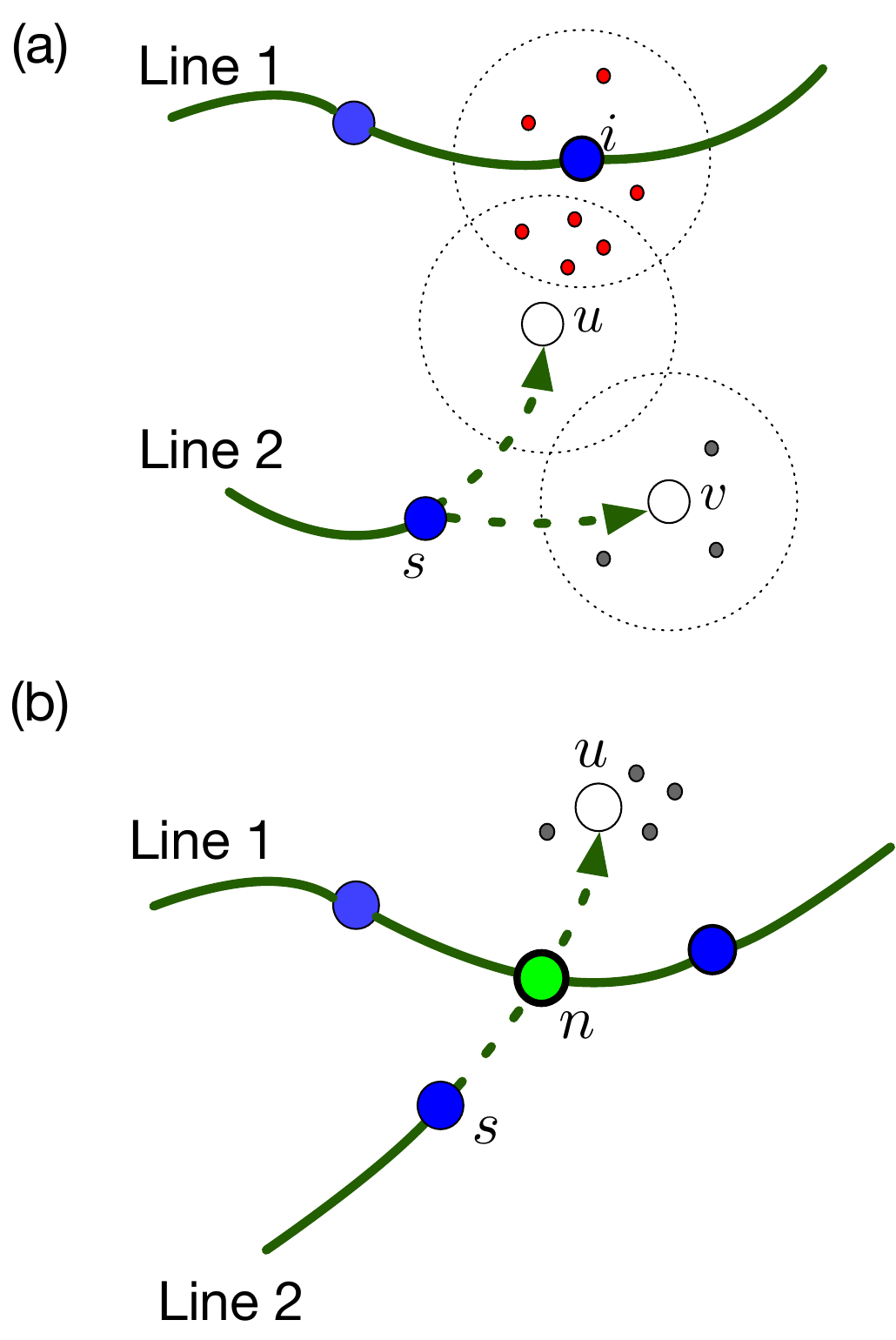} 
    \caption{Illustration of the greedy line growth and intersection mechanism. (a) Node $s$ chooses between candidate locations $u$ and $v$ based on coverage. Points around $u$ (shown in red) have already been covered (by node $i$), giving a coverage $C(u) = 0$, while $v$ has coverage $C(v) = 3$. The location $v$ is selected, inducing an effective repulsion between the two lines. (b) In the case where $s$ grows towards location $u$, the intersections are detected geometrically and converted into network nodes. Here it will give a new node $n$. The resulting network is connected and planar.}
    \label{fig:mecha}
\end{figure}

The network is composed of $M$ lines, each of fixed length $L$ and initially made of $n$ equally spaced nodes. The inter-node distance is $d = L / (n - 1)$. A node covers all points within a disk of radius $d/2$ centered on it. The objective is to maximize the number of previously uncovered points serviced by each newly added node (for an efficient coverage computation, instead of scanning all points, we use a $k$-d tree data structure \cite{bentley1975multidimensional}, which allows efficient spatial queries). Lines are then constructed iteratively in two symmetric directions from a central seed node. For each new segment added (of length $d$), candidate directions are evaluated within an angular sector of opening $2\theta$ centered on the current orientation. Among these candidates, the next node is selected greedily to maximize the number of uncovered points within radius $d/2$. This induces an effective repulsion between lines, as illustrated in Fig.~\ref{fig:mecha}a. For example, the  node $s$ must choose between candidate locations $u$ and $v$. Although $u$ lies closer to a dense region, the surrounding points were already covered (by the node $i$ on a previous line), yielding a number of uncovered points $C(u) = 0$. In contrast, $v$ covers new points (with coverage $C(v) = 3$), making it the preferred choice.

\begin{figure*}[htp]
\centering
   \includegraphics[width=0.95\linewidth]{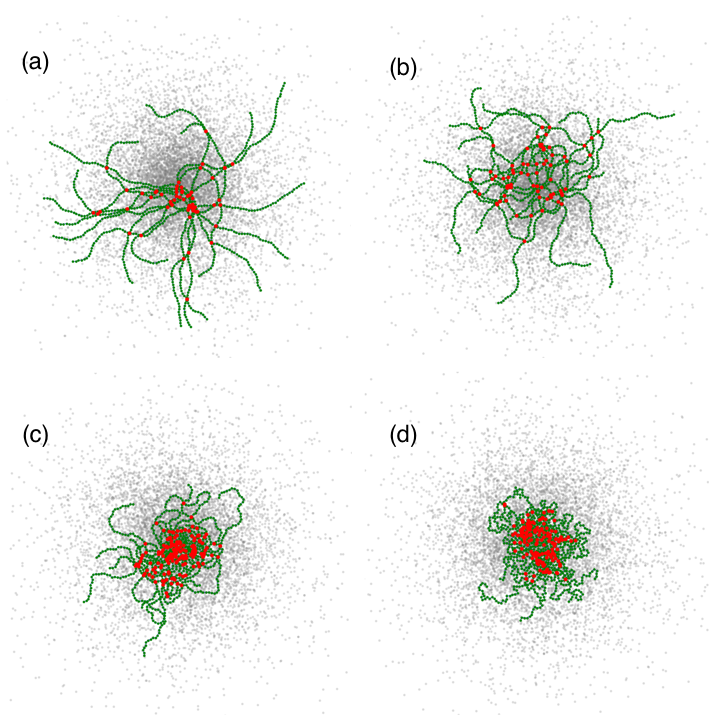}
   \caption{Example of generated spatial networks with different deviation angles (a) $\theta=20^\circ$, (b) $30^\circ$, (c) $45^\circ$ and (d) $90^\circ$. All the networks are obtained for $M = 10$ lines, each composed of $n = 101$ nodes separated by a distance $d = 1$, resulting in lines of length $L = 100$. The network is embedded in a disk of radius $R = 100$ and constructed over $P = 10^4$ sampled points, with density decaying exponentially with scale $r_0 = 10$. Intersections (converted into nodes) are shown in red, highlighting the core and branches structure.}
   \label{fig:example}
\end{figure*}

The network is built sequentially and the first line always starts from the origin. For subsequent lines, a random seed location is selected within the disk, and a candidate line is grown using the greedy procedure described above. To ensure global connectivity, a line is accepted only if it intersects a previously constructed line—either via a shared node or a geometric intersection between edges. When a new line intersects a previously drawn line (without sharing a node) as shown in Fig.~\ref{fig:mecha}b, the intersection point is computed geometrically and explicitly added as a node. The intersecting segments are then split, and the intersection point is connected to the nearest segment endpoints. This transforms a set of independent polylines into a planar network with well-defined topology. Examples of generated networks with different values of $\theta$ (all the rest being equal) are shown in Fig.~\ref{fig:example}.
Variants of the model can be considered, for example, by omitting intersection nodes, which leads to non-planar graphs. We could also explore variants where the growth of shorter lines stops upon intersecting existing structures, reflecting mechanisms described in fungal systems~\cite{lucas2025private,edelstein1982propagation}.

Once the network is generated, we compute various observables to analyze its structural and spatial properties. In the following sections, we examine the scaling behavior of key quantities and compare them with real-world data. We focus on two regimes: (i) how network structure evolves with increasing line length (at fixed number of lines), and (ii) how the number of lines affects structure when line length is fixed.

\section{Scaling with line size}

\paragraph{The ‘Flory’ Exponent.} In polymer physics, the Flory exponent $\nu$ characterizes how the spatial extent of a polymer scales with its length. A common measure of this extent is the radius of  gyration defined as 
\begin{align}
R_g = \left( \frac{1}{n} \sum_{i=1}^n \left\| \vec{r}_i - \vec{r}_{\text{cm}} \right\|^2 \right)^{1/2},
\end{align}
where $\vec{r}_i$ is the position of node $i$, and $\vec{r}_{\text{cm}}$ is the center of mass of the line. For a line composed of $n$ nodes, the radius of gyration $R_g$ scale as
\begin{align}
R_g \sim n^\nu.
\end{align}
Typical values include $\nu = 1/2$ for ideal random walks and $\nu \approx 0.588$ for self-avoiding walks in 3D~\cite{doi1988theory,nienhuis1982exact,de1979scaling}. In our model, each line consists essentially of $n$ equally spaced stations over a total length $L$, and grows under angular constraints and local effective repulsion. As can be seen in Fig.~\ref{fig:example}, we expect the angular deviation $\theta$ to significantly influence the behavior of the lines: small values of $\theta$ constrain the growth to nearly straight paths (yielding $\nu\approx 1$), whereas larger angles allow for more tortuous trajectories, resulting in smaller values of the Flory exponent $\nu$.
\begin{figure}[h!]
\centering
   \includegraphics[width=0.9\linewidth]{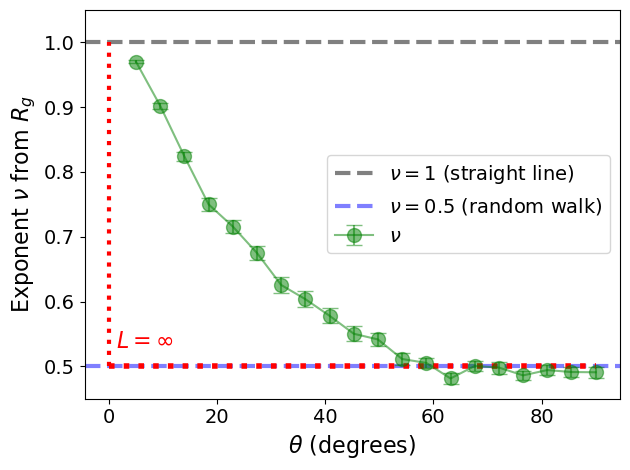}\\
    \includegraphics[width=0.9\linewidth]{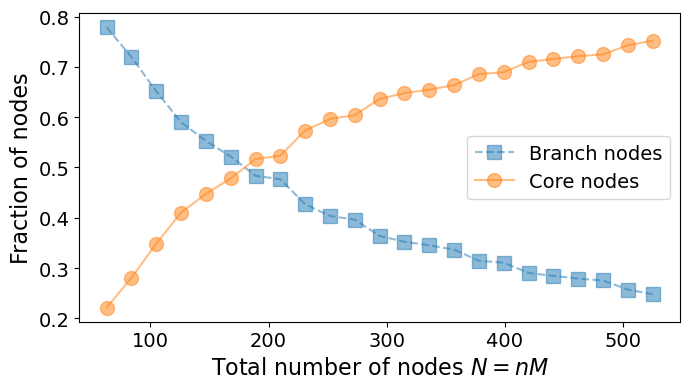}
   \caption{(Top) Evolution of the exponent $\nu$ as a function of the angle $\theta$. For small angles, the behavior is close to a straight line, while for large angles, we recover a random walk behavior with $\nu=1/2$. Results are here obtained for $M=20$ lines, $R=100$, $r_0=10$, $P=10,000$ points, $C=100$ configurations, and the size of lines varying from $L=10$ to $L=200$. In the $L\to\infty$ limit, we expect to recover the correlated walk result \cite{tojo1996correlated},  shown by the red dotted line. (Bottom) Simulated fractions $1-\eta_b$ and $\eta_b$ of the core and branch nodes respectively, for the model when we vary the number of lines $M$ ($R=100$, $r_0=10$, $n=51$, $L=50$, $P=10^4$, $\theta=10^\circ$, $C=100$ configurations). For subway networks with $N\sim 100-300$, this fraction was measured $\eta_b\approx 1/2$ \cite{Roth:2012}.
   }
   \label{fig:nu}
\end{figure}
We show in Fig.~\ref{fig:nu}(top) the measured Flory exponent $\nu$ as a function of the angular deviation $\theta$ for line lengths up to $L = 200$ (see Fig.~A1 in the Appendix for plots of $R_g$ versus $n$ at various $\theta$ values). As expected, for small angular deviations, the trajectories exhibit strong directional persistence, and we observe $\nu \approx 1$, consistent with nearly straight-line motion. For larger deviations, the direction becomes rapidly randomized, and $\nu$ approaches the random walk value $\nu \approx 0.5$ (within error bars). As a result, for finite lengths $L$, we observe an effective exponent in the range $1/2 < \nu < 1$. For very large $L$, however, we expect a different behavior, as can be shown for a correlated random walk with the same angular constraint $\theta$~\cite{tojo1996correlated}. In the limit $L \to \infty$, one finds (see Fig.~A2) $\nu = 1/2$ for any $\theta > 0$ (and $\nu = 1$ for $\theta = 0$), with a crossover length $n^*$ that can be extremely large. In fact, this crossover length diverges as $n^*\sim \theta^{-\mu}$ for $\theta \to 0$ (with $\mu\approx 1.9$ for the correlated random walk \cite{tojo1996correlated}). In most physical systems, however, the infinite-size limit is not relevant, and intermediate or finite-size scaling exponents between $1/2$ and $1$ may be observed in experiments.

\section{The fractal dimension} 

For systems growing from a central point, the fractal dimension $d_f$ can be estimated by counting the total number of nodes $N_{tot}(r)$ within a disk of radius $r$, and varying $r$ \cite{bunde2012fractals}. In general, one expects a scaling relation of the form
\begin{align}
N_{tot}(r) \sim r^{d_f}.
\end{align}
For subway networks, the behavior of $N_{tot}(r)$ was studied in~\cite{benguigui1995fractal,Roth:2012} and displays a uniform density core with $d_f=2$ and a fractal dimension less than one for $r>r_c$ ($r_c$ is the core radius) signalling the one-dimensional nature of branches. For mycelial development, the fractal dimension vary considerably between species \cite{fricker2017}, and has been observed to increase over time, with late-stage values typically in the range $d_f \in [1.6, 1.8]$~\cite{de2015automated,fricker2008interplay} (except for one outlier exhibiting a dimension above 2). The fractal dimension of vascular networks has also been estimated in multiple contexts. In the human retina, the vascular structure typically exhibits also $d_f \in [1.6, 1.8]$~\cite{masters2004fractal}, while bronchial trees in mice show values in the range $d_f \approx 1.54\text{--}1.67$ across different strains~\cite{pmc_bronchial}. These values reflect the space-filling and hierarchical organization characteristic of branching networks in living organisms.

In our model, we compute the fractal dimension over a range of values for $r_0$, using a disk of radius $R = 100$, $P = 10{,}000$ points, line length $L = 100$, and $M = 20$ lines. The results are averaged over values of $r_0$ and angular deviation $\theta$, with 100 independent configurations per setting. We obtained
\begin{align}
d_f = 1.69 \pm 0.18,
\end{align}
and we checked that $d_f$ is not significantly affected by variations in $\theta$ (see Fig.~A3). This value is consistent with those observed in biological systems, such as filamentous mycelial  networks~\cite{jonathan2006reorganization}. This consistency highlights our model’s potential to capture essential features of spatial networks beyond transportation networks.


\section{Scaling with the number of lines}

We now study the effect of interactions between lines on the overall structure of the network. For this we keep the length $L$ of each line fixed (made of $n$ nodes), but we vary the number of lines $M$.\\

\paragraph{Core and branches structure.} We observe in our model the spontaneous emergence of a \emph{core-and-branches} structure. Branches consist of nodes starting from terminal positions ($k=1$), continuing through chains of degree $k=2$ nodes, and eventually merging into the core, which contains nodes with $k > 2$. This structural motif has been observed empirically across the largest subway systems, as documented in~\cite{Roth:2012}, where a key metric was introduced to characterize it: the fraction of nodes in branches, $\eta_b = N_b / N$, where $N_b$ is the number of branch nodes, and $N$ the total number of nodes ($N\approx Mn$ up to the number of intersections points). Empirically, as networks grow larger, the ratio $\eta_b$ tends to approach a value slightly below $1/2$ \cite{Roth:2012}. Using our model, we compute $\eta_b$ and the result is shown in Fig.~\ref{fig:nu}(bottom).
We observe that increasing the number of lines, and thus the overall network size, leads to a larger core. This is because the probability of intersection grows with the number of lines, reducing the relative size of branches while expanding the core. Interestingly, for network sizes comparable to real-world systems ($N \sim 100$--$300$), we find that the proportion of branch nodes stabilizes around $\eta_b \approx 1/2$, consistent with empirical observations reported in~\cite{Roth:2012}.

\section{Number of intersections}

We now consider the number of intersection points with degree $k > 2$, denoted by $N(k > 2)$. These nodes correspond to connection stations in subway networks, cross-links in polymer systems, or anastomoses in filamentous fungi. As such, they capture critical aspects of the network formation process.

In a null model of $M$ randomly oriented line segments, one expects the number of intersections to scale quadratically (for $M\gg 1$) and obeys the following scaling form
\begin{align}
N_{\text{random}}(k>2) \sim \frac{1}{2}M^2 G\left(\frac{L}{2R}\right),
\end{align}
where $G$ encodes the dependence on the line length compared to the total area (see Fig.~A4-A5 and the appendix for an approximate estimate of $G$ for this null model). In our model, the greedy growth mechanism combined with the enforced intersection condition induces an effective repulsion between lines. As a result, the number of intersections scales sub-quadratically:
\begin{align}
N(k > 2) \sim M^\tau,
\end{align}
with $\tau =1.52 \pm 0.13$ (see Figs.~A6-A7), in contrast to the null model of non-interacting lines, where $\tau = 2$.  More precisely, we find
\begin{align}
\tau \approx 1.52 \pm 0.13,
\end{align}
a value that appears largely independent of $r_0$, $L$, or $\theta$ (see Fig.~A7 in the Appendix). This deviation from the trivial exponent confirms that spatial optimization significantly impacts the network structure. The effective repulsion between lines not only reduces the number of intersections but also lowers the scaling exponent itself.

Empirically, we observe a power-law behavior in subway networks with an exponent $\tau \approx 1.22$ (up to $N \approx 250$, see Fig.~\ref{fig:gammas}(a)). Although this value differs from that of the model, it remains well below $\tau = 2$, suggesting the presence of effective repulsive interactions between lines in real systems 
as well.

\section{Scaling of distances}

We now analyze the effect of the number of lines $M$ on the spatial distribution of nodes by computing the average distance from the center for terminal nodes $\overline{d}_T$ (nodes with degree $k=1$), branch nodes $\overline{d}_B$ (chain nodes with $k=2$), and core nodes $\overline{d}_C$. For each node type, the average is taken over all nodes of that type. We find that these distances scale with the number of lines $M$ (at fixed number $n$ of nodes per line) according to power laws of the form
\begin{align}
    d_i\sim M^{\gamma_i},
\end{align}
where $i=t,b,c$ corresponds to terminal, branch, and core nodes, respectively. We show in Fig.~A8 examples of graphs generated with identical parameters, varying only the number of lines from $M = 10$ to $M = 100$. As observed, the average distances of branch, core, and terminal nodes all increase with $M$, reflecting the effective repulsion between lines. The rising probability of intersections with increasing $M$ also leads to an expansion of the core region, pushing branch nodes farther from the center.
\begin{figure}[h!]
\includegraphics[width=0.9\linewidth]{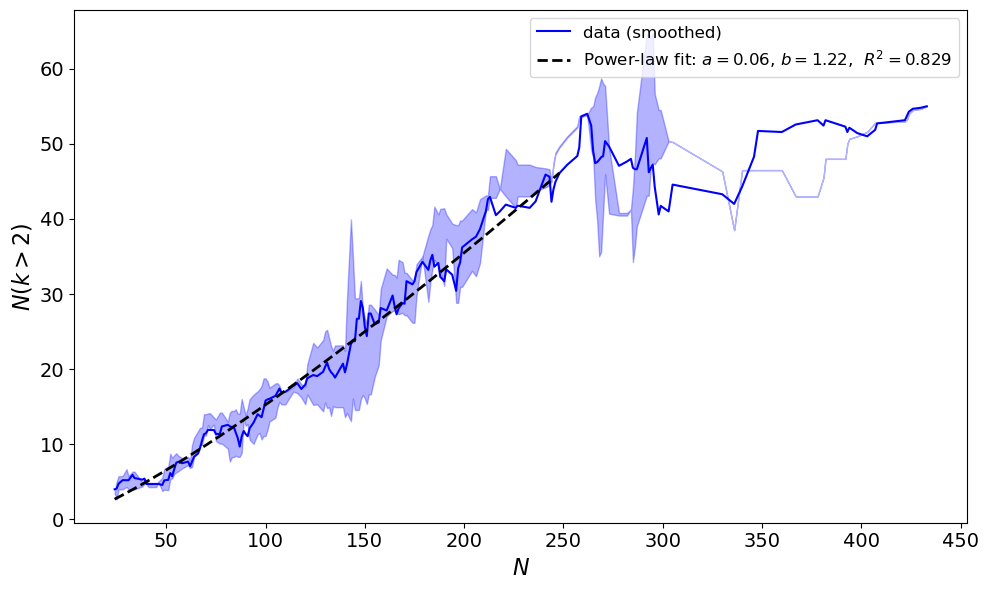}\\
   \includegraphics[width=0.9\linewidth]
   {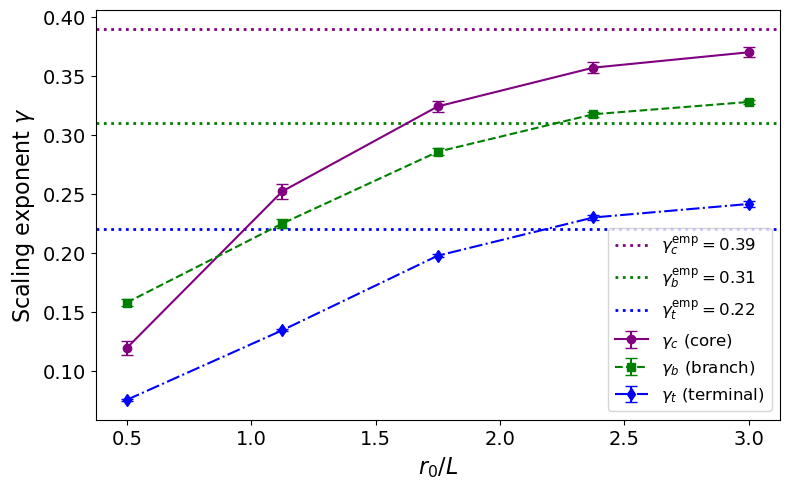}
   \caption{(a) Empirical scaling of the number of connecting nodes (with degree $k > 2$) in subway networks as a function of total network size $N$. A power-law behavior with exponent $\tau \approx 1.22$ is observed up to $N \approx 250$. For larger networks, the small sample size leads to increased noise in the data. (b) Variations of the scaling exponents $\gamma_i$ ($i=t,b,c$) as a function of $r_0 / L$. Horizontal dashed lines indicate empirical reference values derived from the 12 largest subway networks. These results are obtained by varying the number of lines up to $100$ (lines are of size $n=21$, $L=20$, and we used $R=100$, $P=10,000$ points and averaged over $100$ configurations.
   }
   \label{fig:gammas}
\end{figure}

We focus on subway networks as a paradigmatic example to test the predictions of our line-based growth model. Specifically, we analyze the historical evolution of nine of the world’s largest metro systems—London, Madrid, New York City, Mexico City, Moscow, Paris, Seoul, Shanghai, and Tokyo—building on earlier studies of transportation network development~\cite{Xie:2009,Roth:2012}. The empirical scaling exponents associated with the spatial distribution of nodes are approximately $\gamma_t \approx 0.22$, $\gamma_b \approx 0.31$, and $\gamma_c \approx 0.39$ for terminal, branch, and core nodes, respectively (see Fig.~A9). Although there is a large uncertainty about these values, they provide a useful benchmark for validating the model. These scalings imply that the quantity $\eta=\overline{d}_b/\overline{d}_c$ studied in \cite{Roth:2012} is then varying as $N^{\gamma_b-\gamma_c}\approx N^{-0.09}$, which is consistent with the empirical fit that shows a decrease (with a larger exponent but given the noise, we cannot expect a perfect agreement, see Fig. A10). 

We compute these exponents for the model at each value of $r_0/L$ and report the results in Fig.~\ref{fig:gammas}(b). As $r_0$ increases, the point distribution becomes more spatially homogeneous, allowing more points to be found at larger distances from the center. This simple geometric consideration suggests that the average distances of terminal, branch, and core nodes should all increase with $r_0$. The non-trivial observation, however, is that this also alters the corresponding scaling exponents, making them explicitly dependent on $r_0$. In the Appendix (Fig.~A11), we show the individual plots of these distances versus $N$, clearly confirming the variation of the exponents with $r_0$. We further observe that for large values of $r_0/L$, the exponents predicted by the model match those measured in subway networks, underscoring the importance of the network being grown by interacting lines that both repel and intersect.

\section{Perspectives}

Many real-world networks—from transportation systems to biological structures—do not grow by the incremental addition of nodes, but rather through the expansion of extended, line-like elements. This observation suggests a shift in how we model spatial network formation: away from node-based rules and toward frameworks that capture the geometric and topological constraints of growing lines. The model presented here offers such a shift, emphasizing how coverage optimization, angular fluctuations, and intersection conditions can collectively shape complex network architectures.

Beyond its immediate relevance to subway systems, this approach may inform the study of biological filaments, vascularization, or synthetic polymers, where growth, competition, and spatial constraints are central. These systems often exhibit features—core-periphery organization, branching hierarchies, heterogeneous node roles—that may arise generically from line-based growth mechanisms. Establishing whether such features are universal across domains remains a compelling research direction \cite{fricker2009adaptive}.

The model is intentionally minimal, but its possible extensions are manifold. Incorporating density gradients, multiple centers, or obstacles would bring it closer to urban and biological realities. Allowing lines to merge, bifurcate, or follow stochastic paths—rather than a purely greedy coverage logic—could capture a broader range of morphologies. Ultimately, developing a theoretical framework for spatial networks built from lines may bridge disparate systems and offer deeper insight into the geometrical principles underlying real-world infrastructure and living matter.

\section*{Acknowledgements}
MB thanks M. Friker, T. Shimizu for interesting discussions.

\section*{Code availability}
The code that generates the lines and the network is freely available for download at: \url{https://zenodo.org/records/15741969}.


\appendix*
\renewcommand{\thefigure}{A\arabic{figure}}
\setcounter{figure}{0}

\section{Flory exponent $\nu$}

We show on Fig.~\ref{fig:nu} the scaling of $R_g$ for different values of $\theta$.
\begin{figure*}
\centering
\includegraphics[width=0.45\textwidth]{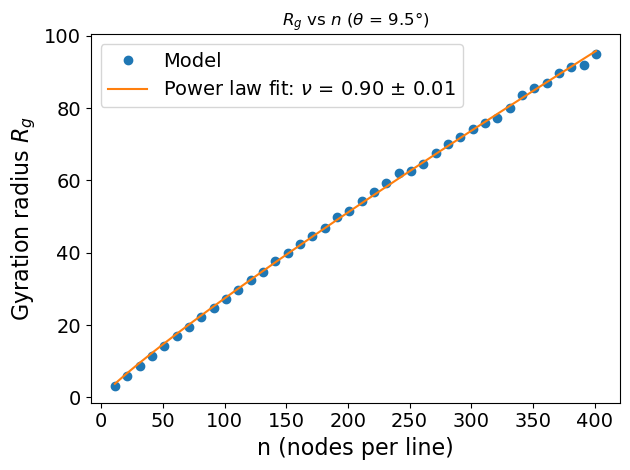}
\includegraphics[width=0.45\textwidth]{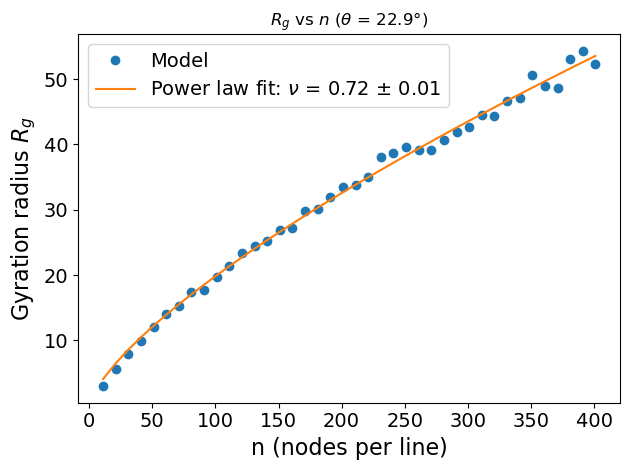}\\
\includegraphics[width=0.45\textwidth]{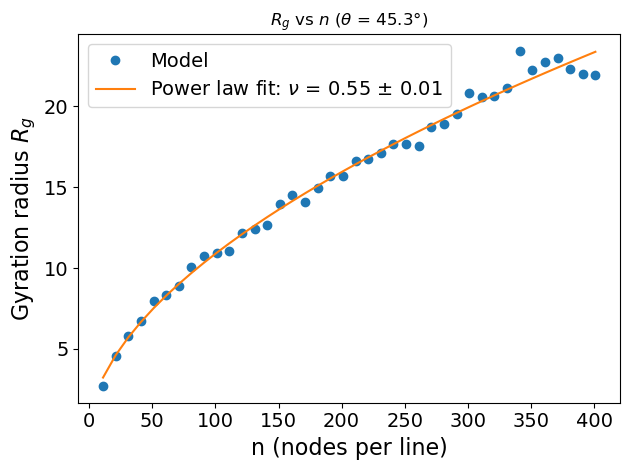}
\includegraphics[width=0.45\textwidth]{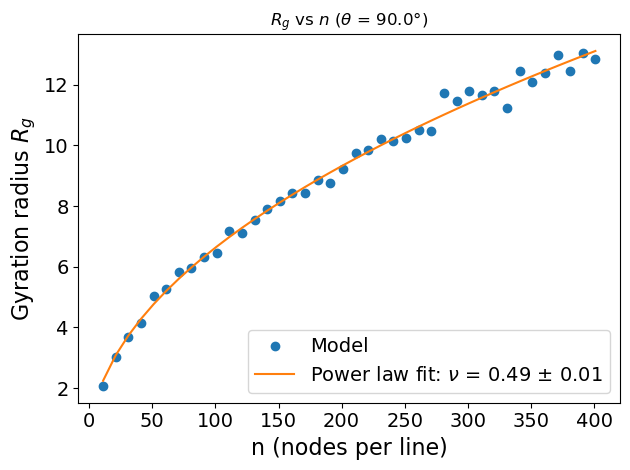}
	\caption{Gyration radius $R_g$ versus $n$ for different values of $\theta$.}
	\label{fig:nu}
\end{figure*}
These results are obtained for relatively small values of $L$ and when $L$ is getting very large, the behavior for $\theta>0$ is the following (see \cite{tojo1996correlated} and references therein)
\begin{align}
R_g^2\sim f(\theta)L
\end{align}
where the prefactor is given here by 
\begin{align}
f(\theta)=\frac{\theta+\sin\theta}{\theta-\sin\theta}
\end{align}
(for $\theta=0$, the walk is a straight line with exponent $\nu=1$).

The crossover to the asymptotic behavior depends strongly on $\theta$ and in \cite{tojo1996correlated}, it was shown that the crossover length $L_*$ (time in \cite{tojo1996correlated}) is behaving as
\begin{align}
L_*\sim \frac{C}{\theta^\mu}
\end{align}
where $\mu\approx 1.9$ and $C\approx 10^5$ (estimated from the Fig. 2 of \cite{tojo1996correlated}). For $\theta$ small, this crossover length can then become extremely large and irrelevant for finite size lines. 

\begin{figure}
\centering
\includegraphics[width=0.4\textwidth]{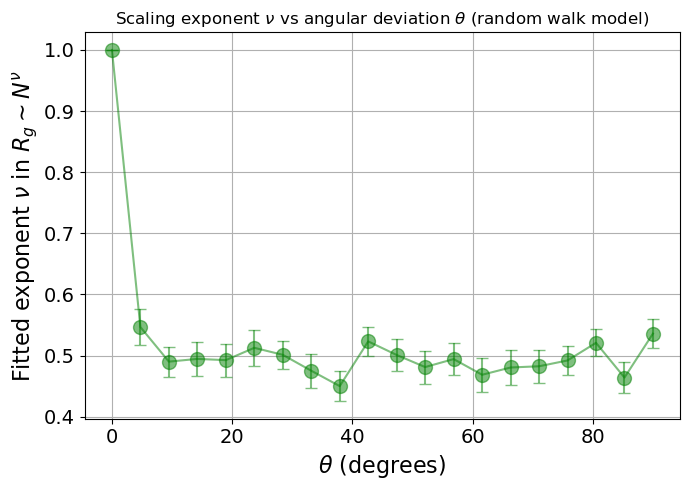}
	\caption{Gyration radius exponent for the correlated random walk model of \cite{tojo1996correlated}. The exponent is obtained by fitting $R_g(L)$ for $L$ up to $10^5$ and averaged over 10 configurations.}
	\label{fig:nu2}
\end{figure}

We plot in Fig.~\ref{fig:nu2} the behavior of $\nu$ versus $\theta$ for correlated random walks (such as in \cite{tojo1996correlated}) which confirms this behavior for large sizes. 

\section{Fractal dimension $d_f$}

\begin{figure}
\centering
\includegraphics[width=0.45\textwidth]{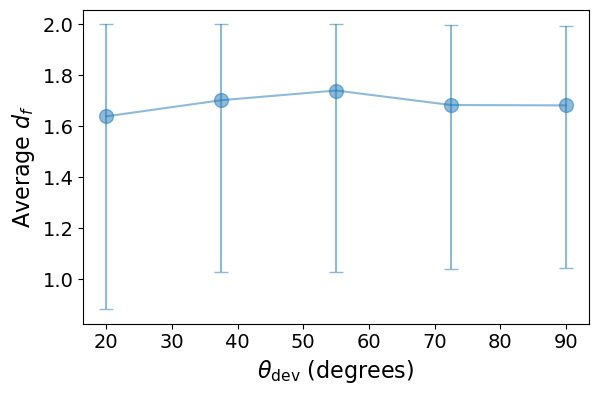}
	\caption{Simulation result for the fractal dimension $d_f$ averaged over $r_0$ for different values of the angle $\theta$.}
	\label{fig:df}
\end{figure}

We study here the impact of the angle $\theta$ on the fractal dimension $d_f$ and the result is shown in Fig.~\ref{fig:df}.
The global mean is $d_f = 1.69$, with a standard deviation of 0.184. This result indicates that the angular deviation 
$\theta$ has no significant impact on the value of $d_f$.

\section{A null model: random segments}

\begin{figure}
\centering
\includegraphics[width=0.45\textwidth]{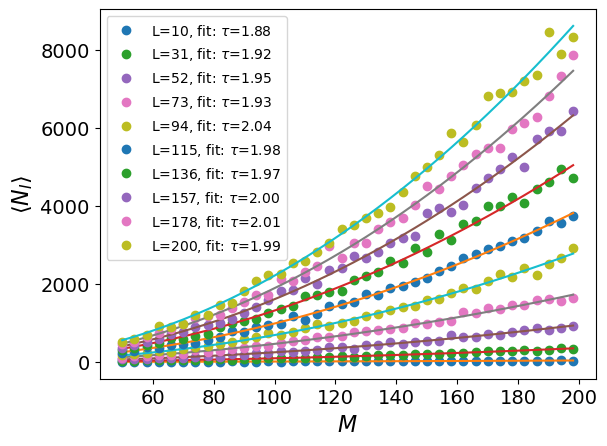}
    \includegraphics[width=0.45\textwidth]{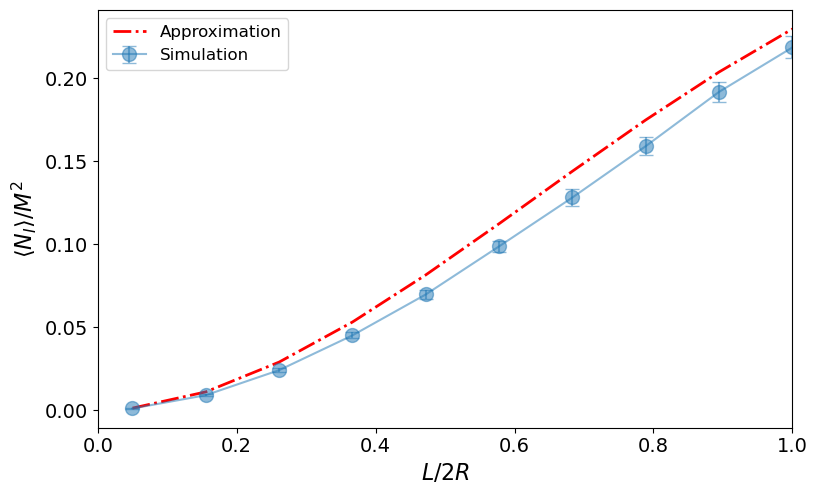}
	\caption{(a) Simulation result for varying $L$ ($R=100$) showing the average number of intersection $\langle N_I\rangle$ versus $M$ for various values of $L$. The power law fit shows that the behavior is always in $M^2$ with a varying prefactor. (b) Comparison of the simulation result with the approximation Eq.~\ref{eq:approx}. The simulation was obtained for $R=100$, $M=200$, and averaged over $100$ configurations. }
	\label{fig:simu}
\end{figure}

We consider $M$ segments of length $L$ whose center is drawn randomly in a disk of radius $R$ (see also \cite{bottcher2020random} where the intersection probability is exactly computed in the case of a unit square). We compute the number $N_I$ of intersections and we show how it varies with $M$ in Fig.~\ref{fig:simu} (we checked that the value of $d$ and hence $L$ doesn't modify the exponent).
We observe on this figure that $N_I$ scales with an exponent $\tau=2$
\begin{align}
N_I\sim M^2
\end{align}
This scaling can be explained as the (maximum) number of intersection is naturally given by $\binom{M}{2}$. We expect the prefactor to depend on $L/R$ and for dimensional reasons to have a function of the form 
\begin{align}
N_I=\frac{1}{2}M^2 G\left(\frac{L}{R}\right)
\end{align}

We provide an approximate estimate of this function using a simple argument. Note that the Parker–Cowan result \cite{parker1976some} (which also recovers Böttcher's calculation \cite{bottcher2020random})
\begin{align}
\frac{N_I}{M^2} = \frac{L^2}{(\pi R)^2},
\end{align}
is expected to hold approximately in the regime $L \ll R$, since it was originally derived for a finite convex body in the plane with uniformly distributed random line segments. We are here sampling a different ensemble than the one assumed by the Parker–Cowan (PC) formula, and the mismatch shows up strongly when $L$ is not small compared to $R$.

Generally speaking, the function $G$ is the probability that two given segments intersect which can be written as
\begin{align}
G\equiv P_{\cap}(L,R)=\int_0^{2R}P_D(x)P_I(x,L)
\end{align}
where $P_D(x)$ is the probability that the centers of the two segments is at distance $x$ and $P_I(x,L)$ is the probability that two segments at a distance $x$ will intersect with each other.

The quantity $P_D(x)$ is a classical problem of geometrical probability and has been computed for any convex body. For a disk of radius $r_0$ is it given by \cite{solomon1978geometric, mathai1999introduction} 
\begin{align}
    P_D(x)=\frac{4x}{\pi R^2}\left[\cos^{-1}\left(\frac{x}{2R}\right)-\frac{x}{2R}\sqrt{1-(x/2R)^2}\right]
\end{align}

\begin{figure}
    \includegraphics[width=0.4\textwidth]{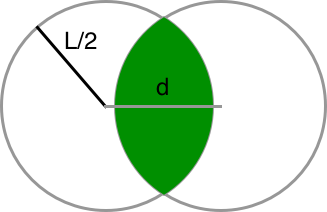}
	\caption{Intersection area (shown in green) of two disks of radius $L/2$ and separated by a distance $d$.}
	\label{fig:diskapprox}
\end{figure}

The probability $P_I(x,L$ that two line segments of length $L$ and with random angles intersect is more difficult to obtain and up to our knowledge has not been exactly computed. A simple approximation consists in replacing it by the area of the intersection of two disk of radius $L/2$ and whose centers are separated by a distance $x$ (see Fig.~\ref{fig:diskapprox}).
This is an upper bound as not all configurations for the segments whose endpoints belonging to this area will give rise to an intersection. The intersection of two disks is a classical problem (see for example \cite{wolfram}) and is given by
\begin{align}
P_I(x,L)=\frac{2}{\pi}\left[\cos^{-1}\left(\frac{x}{L}\right)-\frac{x}{L}\sqrt{1-(x/L)^2}\right]
\end{align}
for $x\leq L$ and $P_I(x,L)=0$ for $x>L$.

Combining these two results, we obtain
\begin{align}
\nonumber
    P_{\cap}(L,R)&=\int_0^{2R}dxP_D(x)P_I(x,L)\\
    \nonumber
    &=\frac{8}{\pi^2R^2}
    \int_0^{\min(2R,L)}xF\left(\frac{x}{2R}\right)F\left(\frac{x}{L}\right)
\end{align}
where
\begin{align}
F(x)=\cos^{-1}(x)-x\sqrt{1-x^2}
\end{align}

We assume that we are in the regime where $L\leq 2R$, and using the variable change $u=x/L$, we obtain
\begin{align}
    P_{\cap}(L,R)=\frac{32}{\pi^2}\left(\frac{L}{2R}\right)^2
    \int_0^1 du u F\left(u\frac{L}{2R}\right)F(u)
\label{eq:approx}
\end{align}

We plot in Fig.~\ref{fig:simu} the results of the simulation together with the numerical integration of the analytical approximation given by Eq.~\ref{eq:approx}, showing that our approximation performs well over the entire range $L \leq 2R$. As expected, the number of intersections increases with the segment length $L$. We also observe that the approximation systematically overestimates the number of intersections.

\section{Intersection exponent $\tau$}


For each value of $r_0$, we plot the $N(k>2)$ versus $N=Mn$ (obtained by varying $M$) and perform a power law fit. In order to estimate the error bar, we compute the value for $\tau_\text{high}$ with a fit over $[N_\text{max}/10, N_\text{max}]$), and we estimate the error as
\begin{align}
\text{error} := \max\left(|\tau_\text{high} - \tau_\text{full}|, |\tau_\text{high} - \tau_\text{low}|\right)
\end{align}
where $\tau_{\text{full}}$ is the fit performed over the whole range $[1,N]$ and $\tau_{\text{low}}$ over range $[1,N_\text{max}/10]$.

To check the dependence on $r_0$ we plot $N(k>2)$ versus $N$ for different values of $r_0$, perform the power law fit and then plot the resulting exponent versus $r_0$. 
\begin{figure}
    \includegraphics[width=0.50\textwidth]{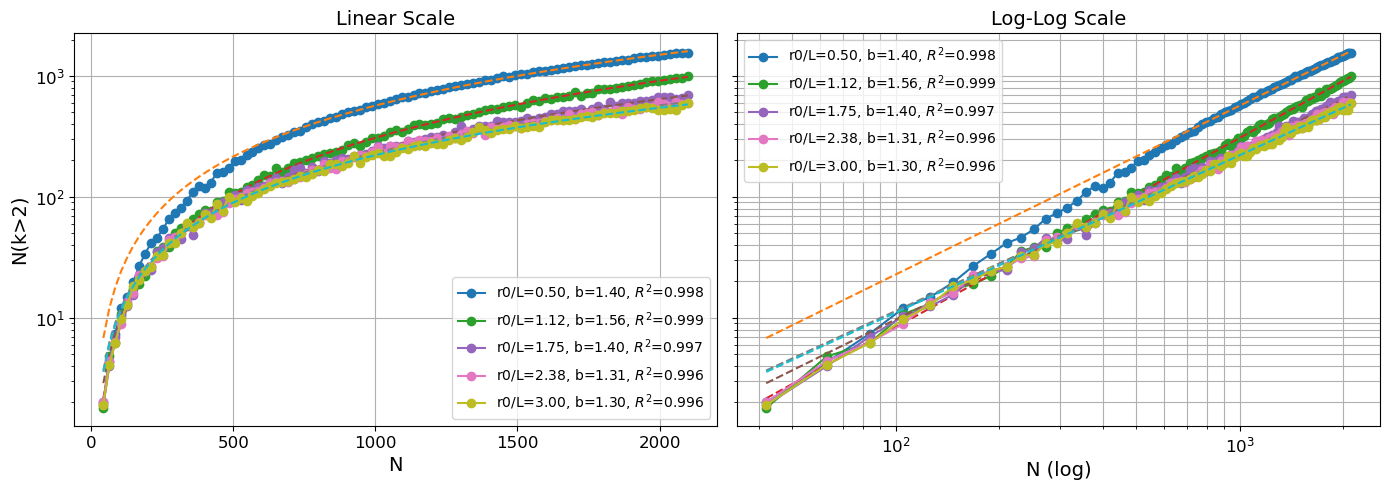}
	\caption{Distances (branches, core, terminal nodes) versus $N$ for different values of $r_0$. }
	\label{fig:taur0}
\end{figure}
In Fig.~\ref{fig:taur0} we show the quantity $N(k>2)$ versus $N$ for different values of $r_0$. In particular, in loglog we indeed cannot observe significant variations of the exponent $\tau$. This is confirmed in the plot of $\tau$ versus $r_0/L$ that gives an average value of $1.40$ with small fluctuations (the std is $\sim 0.1$). These results were obtained for $R = 100$, $n = 21$, $L=20$, $M=100$, $10000$ point, and $C=100$ configurations. 
The result is shown in Fig.~\ref{fig:taur1}(a) 
\begin{figure}
    \includegraphics[width=0.4\textwidth]{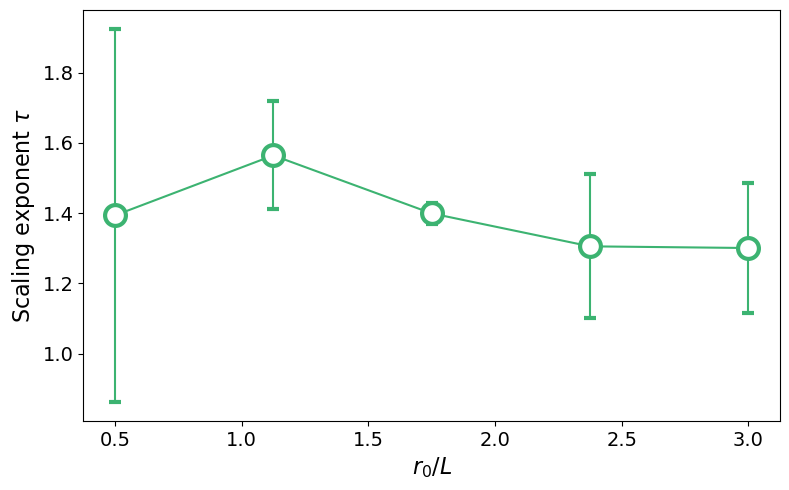}\\
    \includegraphics[width=0.4\textwidth]{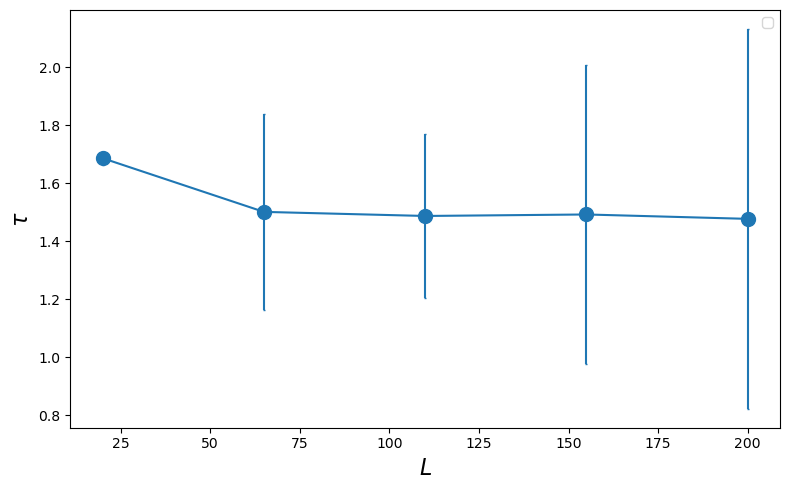}\\
    \includegraphics[width=0.4\textwidth]{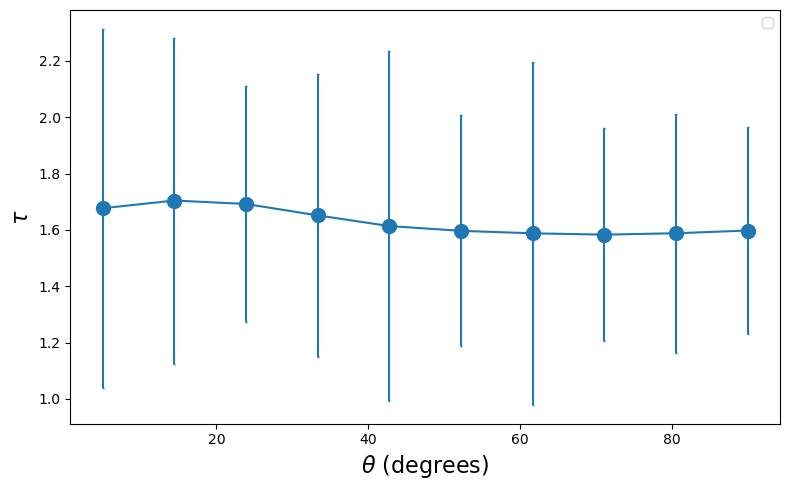}
	\caption{(a) Intersection exponent versus $r_0/L$. (b) Intersection exponent versus $L/2R$. (c) Intersection exponent versus $\theta$. These results are obtained for $R = 100$, $10000$ points, maximum number of lines $M=20$, $r_0 = 10$, $L=10$, $n = 21$, and $C = 100$ configurations.}
	\label{fig:taur1}
\end{figure}
with average value $1.40$ and standard deviation $0.1$.

We also observe that $L$ has no significative impact on the value of $\tau$ and we get $\tau\approx 1.53\pm 0.08$ (Fig.~\ref{fig:taur1}(b)). The exponent $\tau$ doesn't show any significative dependence on $\theta$. In Fig.~\ref{fig:taur1}(c), we show this exponent versus $\theta$. Fluctuations are small and the average is $\tau=1.63\pm 0.05$ (the error bar is given by the std). Averaging all these values gives us
\begin{align}
     \tau \approx 1.52 \pm 0.13,
\end{align}

\section{Scaling of distances}

\subsection{Effect of $M$ on the shape of networks}

\begin{figure*}
    \includegraphics[width=0.4\textwidth]{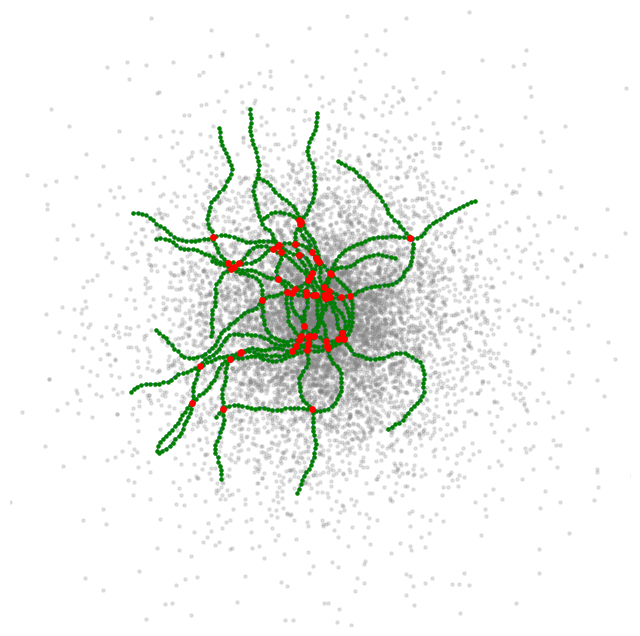}
    \includegraphics[width=0.4\textwidth]{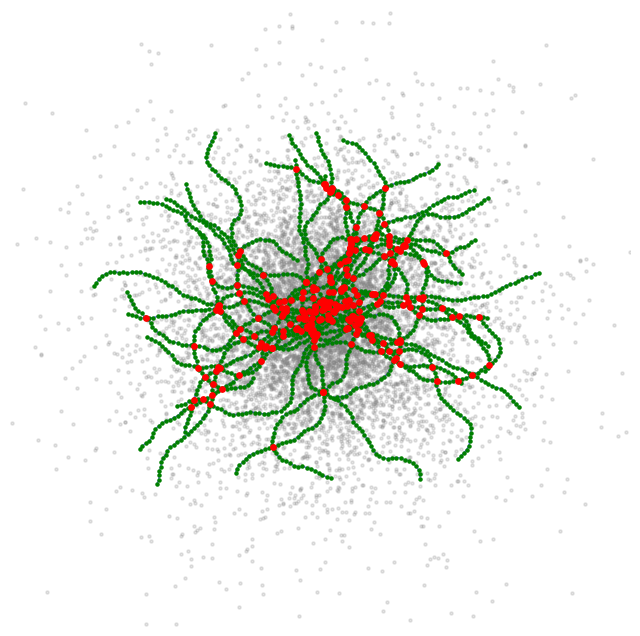}\\
    \includegraphics[width=0.4\textwidth]{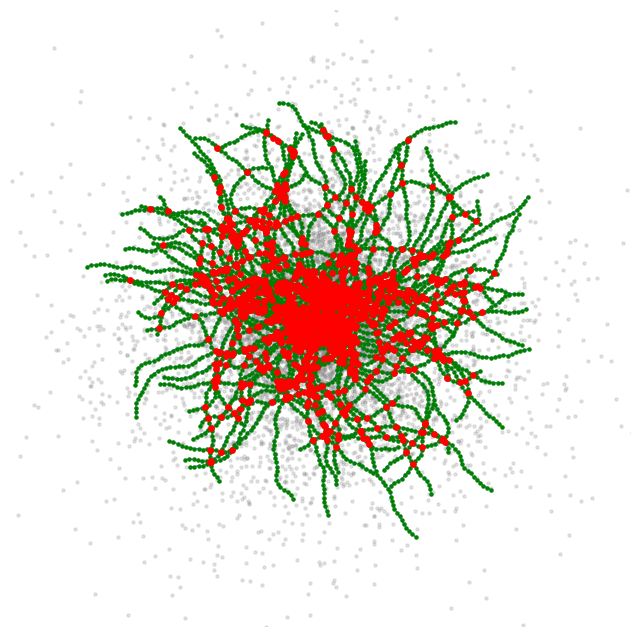}
    \includegraphics[width=0.4\textwidth]{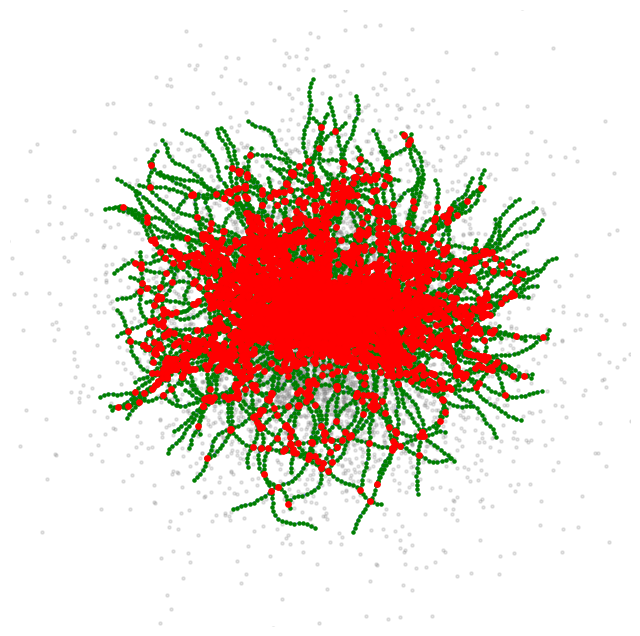}
	\caption{From top to bottom and left to right: graphs obtained for different values of the number of line $M=10, 20, 50, 100$ ($R =100$, $r_0= 10$, $n = 101$, $10,000$ points, $\theta=20^\circ$). The nodes are indicated in red.}
	\label{fig:mm}
\end{figure*}

We show on Fig.~\ref{fig:mm} examples of graphs generated with identical parameters, varying only the number of lines from $M = 10$ to $M = 100$. As expected, the average distances of branch, core, and terminal nodes all increase with $M$, reflecting the effective repulsion between lines. This line repulsion also leads to an expansion of the core region, pushing branch nodes farther from the center.

\subsection{Empirical measure for subways}

\begin{figure}
    \includegraphics[width=0.5\textwidth]{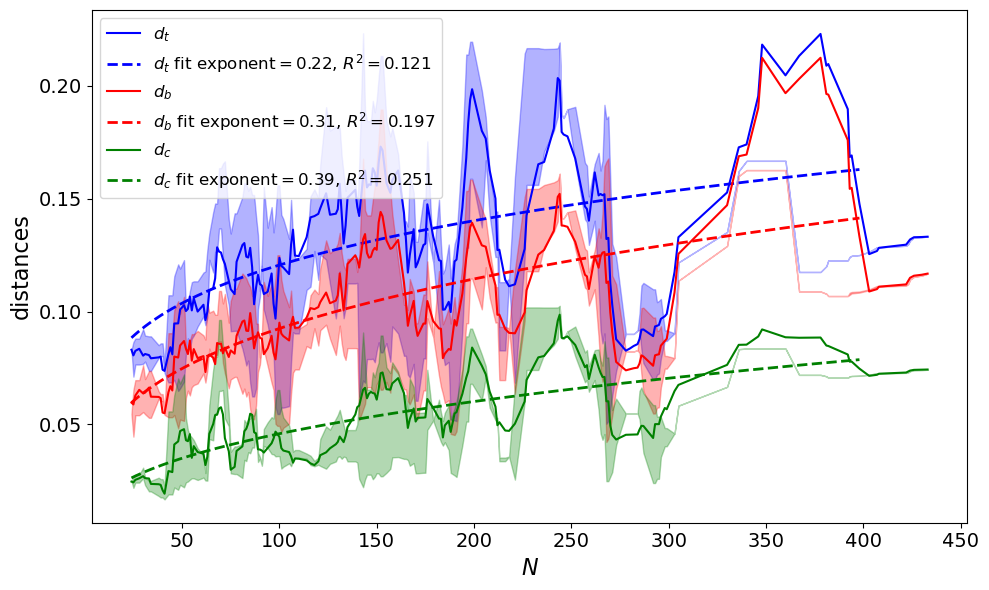}
	\caption{Distances (branches, core, terminal nodes) versus $N$ for the 12 largest subways. }
	\label{fig:emp}
\end{figure}

We show in the Fig.~\ref{fig:emp} the variation of the distances for the core, branches and terminal nodes versus the number of stations. 
We observe distinct exponent values: $\gamma_c^{\text{emp}} = 0.39$, $\gamma_b^{\text{emp}} = 0.31$, and $\gamma_t^{\text{emp}} = 0.22$, although the data exhibit substantial noise.

\begin{figure}
    \includegraphics[width=0.5\textwidth]{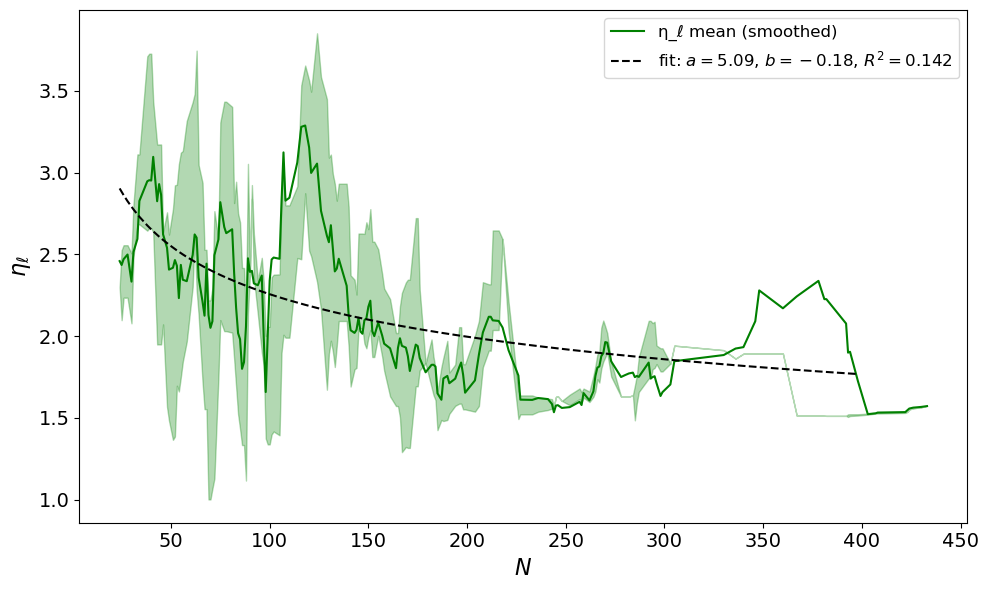}
	\caption{Ratio $\eta_l$ for the subway networks. The power law fit is shown. }
	\label{fig:emp2}
\end{figure}

According to these results the ratio $\eta_l=\overline{d}_b/\overline{d}_c$ computed in \cite{Roth:2012} should behave as
\begin{align}
\eta_l\sim N^{\gamma_b-\gamma_c}
\end{align}
with $\gamma_b-\gamma_c\approx -0.1$.
We observe empirically the result shown in Fig.~\ref{fig:emp2}. 
We observe indeed a decaying function and power law fit gives a value of $\approx -0.18$ but with a small confidence due to the large noise ($r^2=0.14$). This result is however consistent with the model.

\subsection{Scaling of distances for the model}

\begin{figure}
    \includegraphics[width=0.4\textwidth]{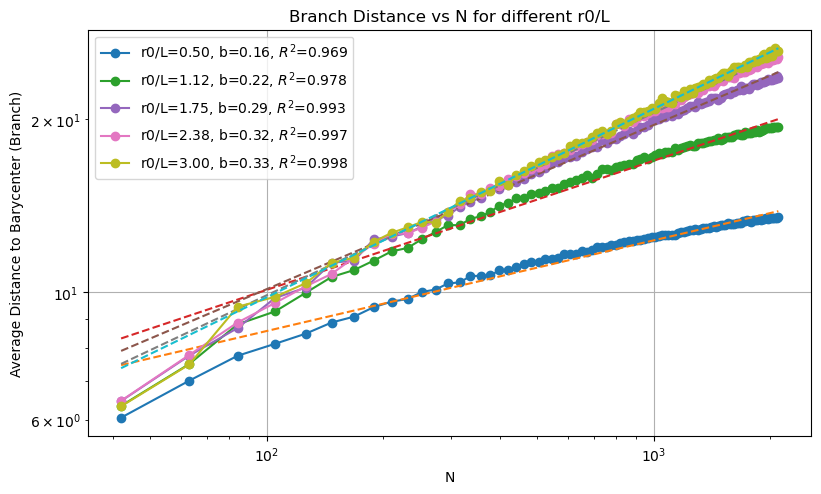}
    \includegraphics[width=0.4\textwidth]{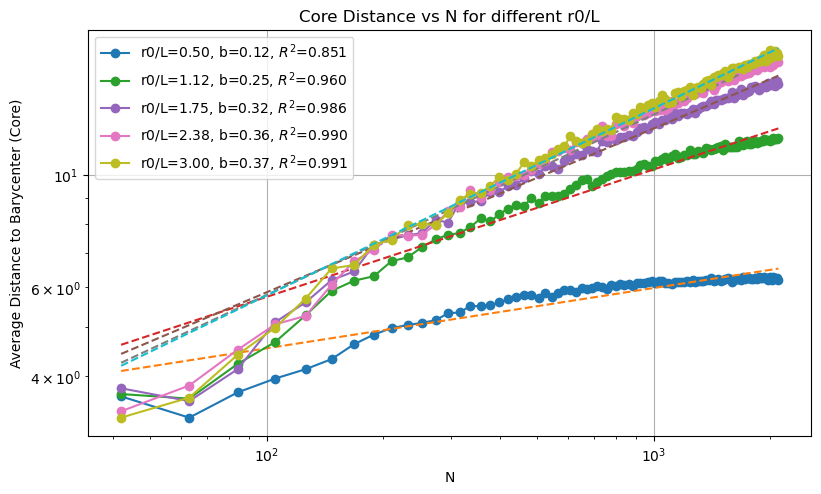}
    \includegraphics[width=0.4\textwidth]{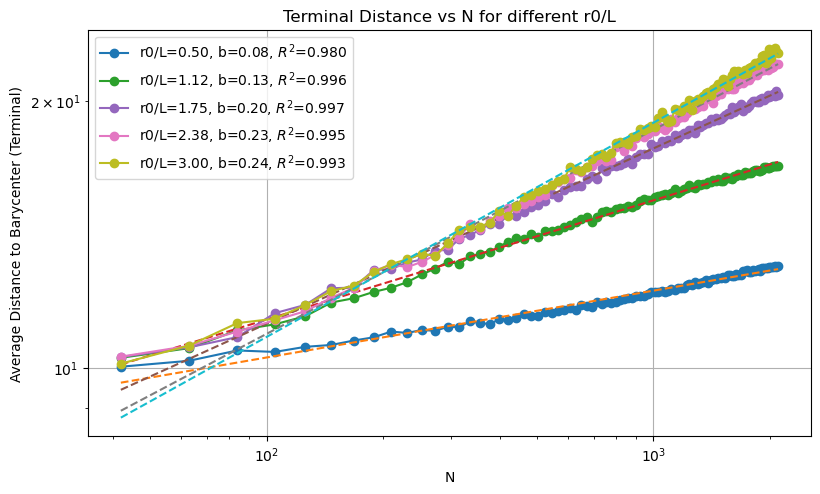}
	\caption{Distances (branches, core, terminal nodes) versus $N=Mn$ (where the number of nodes per line is fixed) for different values of $r_0$.}
	\label{fig:scal}
\end{figure}

We show on Fig.~\ref{fig:scal} the scaling of the distances (for the core, terminal, and branches) for different $r_0$ and that clearly display the variations of the exponents $\gamma_i$. These results were obtained for $R = 100$, $n = 21$, $L = 20$, number of lines maximum $M= 100$, number of points $10000$, $\theta=10$, and $C=10$ configurations (here we vary the number of lines at fixed $n$). 

\clearpage
\newpage

\bibliographystyle{unsrt}
\bibliography{bibfile_network}

\end{document}